RESEARCH ARTICLE

# Intraday Seasonalities and Nonstationarity of Trading Volume in Financial Markets: Individual and Cross-Sectional Features

Michelle B. Graczyk[1], Sílvio M. Duarte Queirós[1,2]*

1 Centro Brasileiro de Pesquisas Físicas, Rio de Janeiro, RJ, Brazil, 2 National Institute of Science and Technology for Complex Systems, Rio de Janeiro, RJ, Brazil

* sdqueiro@google.com

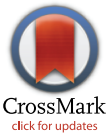

**Citation:** Graczyk MB, Duarte Queirós SM (2016) Intraday Seasonalities and Nonstationarity of Trading Volume in Financial Markets: Individual and Cross-Sectional Features. PLoS ONE 11(11): e0165057. doi:10.1371/journal.pone.0165057

**Editor:** Enrico Scalas, University of Sussex, UNITED KINGDOM

**Received:** May 7, 2016

**Accepted:** September 9, 2016

**Published:** November 3, 2016

**Data Availability Statement:** Data are available from the Chair of Quantitative Finance of CentraleSupélec (Paris, France). Any request on these data must be addressed to the head, Prof Frédéric Abergel (frederic.abergel@ecp.fr). Regarding their availability he has sent us the following confirmation: "The Chair of Quantitative Finance at CentraleSupélec will grant online access to any scientist interested in order book data".

**Funding:** This work was funded by Conselho Nacional de Desenvolvimento Científico e Tecnológico (Dr Sílvio M. Duarte Queirós) and Fundação Carlos Chagas Filho de Amparo à

## Abstract

We study the intraday behaviour of the statistical moments of the trading volume of the blue chip equities that composed the Dow Jones Industrial Average index between 2003 and 2014. By splitting that time interval into semesters, we provide a quantitative account of the nonstationary nature of the intraday statistical properties as well. Explicitly, we prove the well-known ∪-shape exhibited by the average trading volume—as well as the volatility of the price fluctuations—experienced a significant change from 2008 (the year of the "sub-prime" financial crisis) onwards. That has resulted in a faster relaxation after the market opening and relates to a consistent decrease in the convexity of the average trading volume intraday profile. Simultaneously, the last part of the session has become steeper as well, a modification that is likely to have been triggered by the new short-selling rules that were introduced in 2007 by the Securities and Exchange Commission. The combination of both results reveals that the ∪ has been turning into a ⊔. Additionally, the analysis of higher-order cumulants—namely the skewness and the kurtosis—shows that the morning and the afternoon parts of the trading session are each clearly associated with different statistical features and hence dynamical rules. Concretely, we claim that the large initial trading volume is due to wayward stocks whereas the large volume during the last part of the session hinges on a cohesive increase of the trading volume. That dissimilarity between the two parts of the trading session is stressed in periods of higher uproar in the market.

## Introduction

Despite the fact that the state of a financial asset—and the indices composed thereof—is traditionally characterised by the respective price, $S(t)$, and its (percentual) variation, $r_{\Delta t}(t)$, with respect to a given period of time, $\Delta t$, it is nowadays well-established that the complex nature of a financial system cannot be satisfactorily described by a reduced number of quantities [1–4]. For this reason, as we move up from mainstream mass media towards financial platforms we are by default told about quantities like the volatility, $\Sigma$, and the amount of the asset (stocks in our case) that changed hands—i.e., the *trading volume*, $v$—as well as figures about the reference price and the distance to support and resistance levels among other quantities. Owing to the fact that financial markets are part of our social system, in a second horizon, both the volatility and the traded volume provide a means of quantifying the sentiment of the market participants, whose actions depend on the macroscopic market state in a nontrivial way. In other words, if we exclude episodes such as the payment of dividends, (reverse) stock splits and exogenous effects,—for instance, companies that deal with commodities have their price strongly correlated with the product their business is based on, e.g., Middle East political crises have consistently affected crude-oil prices and the value of oil corporations, whatever their nationality —the price of an equity changes when a buyer and a seller agree to trade a quantity $v$ of that asset for a value $S$ that is different from its present quote.

The willingness to buy or sell is mainly based on the assessment about the (under)overpricing of the stock and its expected evolution. That evaluation can be set up by analysing the current and the future economical scenarios, including the fundamentals of the company or its activity sector—just to cite a couple of them —, which is an exercise that obviously involves the handling of information [5]. Therefore, every time a given volume $v$ is traded, information—even if underlying or wrapped as a perception—propagates throughout microscopic constituents of the system, i.e., the financial agents, and eventually fuels the emergence of herding phenomena [6, 7].

The way the flow of information befalls and affects the price dynamics separates out two of the most important theories in finance: the Mixture of Distributions Hypothesis [8] and the Sequential Arrival of Information Hypothesis [9]; the former relates the volatility and the trading volume to information and latent events in financial markets whereas the latter implies that a given amount of information fed into the market will propagate in such a way that it will ignite a series of local steady states until equilibrium is reached at last. In addition, if we bear in mind situations of liquidity—typical of American stock markets—for which the order books are tight, resilient and not very deep, a large trading volume is deemed necessary to stoke a large price variation. That said, we understand the emergence of relevant information tends to materialise in large trading volumes, which create pressure on the order book on its own that at the end triggers large price fluctuations. Although recent studies [10] argue the main origin of large price fluctuations is the emergence of imbalances in the order book instead of large values of $v$, the reality is that in excess of 70% of the daily price fluctuations in blue chip american equities above 4%—which correspond to 4-odd standard deviations—coincide with trading volumes at least two times as large as its daily average, an observation that supports the famous watchword "it takes volume to make the price move".

The interest in a quantitative description of trading volume—or the traded value which combines $v$ and $S$ by means of their product [11, 12]—is old [7]. Among the so-called stylised facts of either quantity, we can highlight the long-lasting autocorrelation function and the fat-tailed stationary distribution, which was initially associated with the log-Normal. From the analysis of high-frequency data, it was found the trading volume distribution, $p(v)$, is actually compatible with asymptotic power-law distributions, namely the $F$-distribution [13–19]. In respect of the autocorrelation function, although prior analyses had pointed to a power-law decay as well [20], subsequent results indicated that quantity is best described by a composition of exponential regimes [21], a result which upheld a "superstatistical" [22] approach to $p(v)$ (for a recent review on trading volume please consult [23]). Combining those results, it is possible to attribute those statistical features to the nonstationarity of the trading activity—defined as the number of active agents in the market—on which the (local) average trading volume depends. That sort of nonstationarity is even verified after treating the data by taking into consideration another well established fact: both the opening and the closing of the trading sessions are identified with markedly large trading volumes resulting in an intraday profile that is widely known as the ∪-shape of financial markets [24, 25]; similar profiles are exhibited by the absolute value of the price fluctuation as well as other forms of computing the volatility [26]. A couple of reasons can be presented to explain this phenomenon: on the one hand, when the market opens, we have the transferring of news and events that came up overnight to the stock price taking into consideration their likely impact on the value of the company, economical sector or country. On the other hand, we have the action of intraday traders who do not carry open positions from one day to the other, *i.e.*, they start and finish the business day with a portfolio entirely composed of cash. Furthermore, the intraday profile of trading can be influenced by idiosyncrasies of the market, namely rulings over the span given to an agent to close a position or set off a derivative.

Recently, the analysis of intraday seasonalities was extended to the higher-order moments of the price fluctuations considering individual and collective statistics [27]. That work showed the existence of interesting intraday profiles in those quantities as well as relations between moments of different order. Taking into consideration that work and the relevance of trading volume in the characterisation of the dynamics of financial markets, in this manuscript, we introduce a comparable analysis of the intraday properties of *v*. Moreover, since a financial market is a well founded representation of an economy—which is known to consist of cycles of contraction and expansion—we go farther afield by also analysing to what extent the aforementioned intraday features have evolved in recent years (assuming a semester-by-semester basis). For the sake of conciseness, we split our work into two parts: analysing the statistical properties related to the first four order cumulants of the trading volume individually and cross-sectionally. In a subsequent work [28], we will discuss the collective intraday features of our data as well as its nonstationarity.

## Materials and Methods

### Data

Our results are obtained using 1-minute frequency data of price and trading volume of the 30 companies constituting the Dow Jones Industrial Average spanning the period between the 4th January 2004 and the 30th December 2013 provided by Olsen Financial Data and the Chair of Quantitative Finance of the École CentraleSupélec. Specifically, the former furnished the data of the second semester of 2004 which we used for a first analysis and the former supplied the data corresponding to the entire set from which the full results are obtained. The companies are their ticker symbols are listed in Table 1 where the suffix ".N" signals the equity is traded at NYSE and ".OQ" at NASDAQ, respectively. Both markets open at 9:30 and close 16:00 and have pre- and post-market periods. We use the following notation: $v_i(d, t; s)$ represents, the 1-minute trading volume of company, $i$, at the intraday time, $t$—is defined in a integer number format with $t = 0$ representing 9:30 in clock time and $t = 390$ corresponding to 16:00 —, on the day, $d$; furthermore, we take into account the semester, $s$, to which $d$ belongs because we divide our data into contiguous semesters in order to assess the nonstationarity of the intraday properties. Such segmentation yields a good balance between quasi-stationarity and a statistically significant number of days within each span so that a higher order statistical analysis can be implemented. The first semester of 2004 (1S04) corresponds to $s = 1$, the second semester of 2004 (2S04) to $s = 2$ and so forth until the second semester of 2013 (2S13), i.e., $s = 19$.

### Main formulae and definitions

Let us consider a general quantity, $\mathcal{O}$, to introduce the notation we will employ hereinafter. Typically, we have $\mathcal{O} = v^n$ ($n = 1,2$), but it can represent the cumulants of the trading volume

Table 1. Companies analysed and their ticker symbols.

| Company | Ticker symbol |
|---|---|
| Alcoa Inc. | AA.N |
| American International Group Inc. | AIG.N |
| American Express Co. | AXP.N |
| Boeing Co. | BA |
| Citigroup Inc. | C.N |
| Catterpilar Inc. | CAT.N |
| E.I. DuPont de Nemours & Co. | DD.N |
| Walt Disney Co. | DIS.N |
| General Eletric Co. | GE.N |
| General Motors Corp. | GM.N |
| Home Depot Inc. | HD.N |
| Honeywell International Inc. | HO.N |
| HewlettPackard Co. | HPQ.N |
| International Business Machines Corp. | IBM.N |
| Intel Corp. | INTC.OQ |
| Johnson & Johnson | JNJ.N |
| JPMorgan Case & Co. | JPM.N |
| Coca-Cola Co. | KO.N |
| McDonald's Corp. | MCD.N |
| 3M Co. | MMM.N |
| Altria Group Inc. | MO.N |
| Merc & Co. | MRK.N |
| Microsoft Corp. | MSFT.OQ |
| Pfizer Inc. | PFE.N |
| Procter & Gamble Co. | PG.N |
| SBC Communications Inc. -> AT&T Inc. | SBC.N -> T.N |
| United Technologies Corp. | UTX.N |
| Verizon Communications Inc. | VZ.N |
| Wal-Mart Stores Inc. | WMT.N |
| Exxon Mobil Corp. | XOM.N |

doi:10.1371/journal.pone.0165057.t001

as well, particularly in the nonstationarity analysis case. That said, we define two kinds of averages; explicitly, we have averages computed over days

$$\overline{\mathcal{O}_i(t;s)} \equiv \frac{1}{N_D} \sum_{d=d_1}^{d_l} \mathcal{O}_i(d,t;s), \qquad (1)$$

where $d_1(d_l)$ corresponds to the first(last) day of the semester $s$ that contains $N_D$ business days with a quote at time $t$ for company $i$. Accordingly, for the first(second) semester of a given year, $d_1$ typically represents the first business day in January(July) and $d_l$ the last business day in June(December). On average, we have 126 days per semester. We could have used other segmentation approaches, e.g., the method introduced in [29] that was already applied in the analysis of financial quantities [30]; however, it is likely that its application would give patches of uneven duration and thus the existence of rather small intervals yielding insufficient statistics for them.

Alternatively, we can average over companies as well. These are defined as

$$\langle \mathcal{O}(d, t; s) \rangle \equiv \frac{1}{N_C} \sum_i \mathcal{O}_i(d, t; s), \quad (2)$$

where $N_C$ represents the number of companies. In our case, $N_C$ is equal to 30 for most of the semesters except for the cases $s$ = 11, 12 and 13, where we consider 29 companies due to the filling by GM of a reorganisation process ("Chapter 11") to the US Bankruptcy Court.

In addition, we can also combine averages over days and companies,

$$\tilde{\mathcal{O}}(t; s) \equiv \langle \overline{\mathcal{O}_i(d, t; s)} \rangle, \quad (3)$$

or the other way round, where averages over companies followed by averages over days,

$$\hat{\mathcal{O}}(t; s) \equiv \overline{\langle \mathcal{O}_i(d, t; s) \rangle}, \quad (4)$$

which we relate to cross-sectional statistical studies. In each case—individual, Eq (1), and cross-sectional, Eq (2)—we denote the average trading volume by $\mu$,

$$\mu_i(t; s) \equiv \overline{v_i(d, t; s)}, \qquad \mu(d, t) \equiv \langle v_i(d, t; s) \rangle, \quad (5)$$

It is straightforward to verify that $\tilde{\mu}(t; s) = \hat{\mu}(t; s)$.

Equivalently, the median—the value lying at the midpoint of the sorted trading volume considering all the values of company $i$ registered at time $t$ for semester $s$ (individual analysis) or the values of all companies at a given time $t$ and day $d$ (cross-sectional analysis)—is represented by either $m_i(t; s)$ or $m(d, t; s)$, respectively.

The quantities we want to analyse, namely higher order statistical moments, are prone to strong error if we directly apply the formulae in Eqs (1) and (2) with $\mathcal{O} = v^n$ ($n \geq 3$) to compute them. For this reason, and aiming at maintaining some coherence between our calculations and the results for the price fluctuations [27], we use the relations that emerge from the Gram-Charlier expansion of a probability density function (PDF) assuming the Normal as the reference distribution [31] to obtain the following definitions for the variance, $\sigma^2$, skewness, $\zeta$, and the kurtosis, $\kappa$,

$$\sigma_i^2(t; s) \equiv \overline{v_i^2(d, t; s)} - \mu_i(t; s)^2, \quad \sigma^2(d, t; s) \equiv \langle v_i^2(d, t; s) \rangle - \mu(d, t; s)^2, \quad (6)$$

$$\zeta_i(t; s) \equiv \frac{6}{\sigma_i(t; s)} [\mu_i(t; s) - m_i(t; s)], \quad \zeta(d, t; s) \equiv \frac{6}{\sigma(d, t; s)} [\mu(d, t; s) - m(d, t; s)], \quad (7)$$

$$\kappa_i(t; s) \equiv 24 \left( 1 - \sqrt{\frac{\pi}{2}} \frac{\overline{|v_i(d, t; s) - \mu_i(t; s)|}}{\sigma_i(t; s)} \right) + \zeta_i(t; s)^2,$$

$$\kappa(d, t; s) \equiv 24 \left( 1 - \sqrt{\frac{\pi}{2}} \frac{\langle |v_i(d, t; s) - \mu(d, t; s)| \rangle}{\sigma(d, t; s)} \right) + \zeta(d, t; s)^2. \quad (8)$$

In employing Eqs (7) and (8) we avoid calculating further statistical moments other than the average, the variance as well as the median and the significant addition error.

Consonantly, the hat and the tilde notation extends to the cumulants of $n$-th order, $\mathcal{K}_n$, given in Eqs (5)–(8), so that $\widetilde{\mathcal{K}}_n$ means the cumulant is computed first considering a statistics over days and then the results are averaged over companies whereas $\widehat{\mathcal{K}}_n$ means the cumulant is computed using a statistics over companies which is afterwards averaged over days.

### Welch's `t`-Student test for equivalence of the means

The Welch's `t`-Student test for the equivalence of the means is based on the Student's `t`-distribution with $n$ degrees of freedom

$$p(\mathtt{t}) \propto \left(1 + \frac{\mathtt{t}^2}{n}\right)^{-\frac{n+1}{2}},$$

where it is imposed the null hypothesis that the mean of two samples are equal whereas the alternative hypothesis assumes that they are different, i.e., two samples belong to different and independent populations. In our case, the two groups consist of the values of the opening trading volume relaxation exponent $\alpha$ from 1S04 until 1S08 (yielding $n_1 = 9$ elements to the first group) and from 2S08 until 2S13, which has $n_2 = 10$ elements. For these two sets, we compute the average and the variance of both, which are equal to $\boldsymbol{\alpha}_1 = 0.29$ and $\boldsymbol{\sigma}^2_{\boldsymbol{\alpha}_1} = 1.09 \times 10^{-4}$ for the first set and $\boldsymbol{\alpha}_1 = 0.37$ and $\boldsymbol{\sigma}^2_{\boldsymbol{\alpha}_1} = 1.11 \times 10^{-3}$ for the second set of values. In this Welch's test, we have $\mathtt{t} = \boldsymbol{\alpha}_1 - \boldsymbol{\alpha}_2$ and

$$\boldsymbol{\sigma}^2_{\boldsymbol{\alpha}} = \frac{\boldsymbol{\sigma}^2_{\boldsymbol{\alpha}_1}}{n_1} + \frac{\boldsymbol{\sigma}^2_{\boldsymbol{\alpha}_2}}{n_2}$$

whereas the number of degrees of freedom reads

$$n = \frac{\boldsymbol{\sigma}^4_{\boldsymbol{\alpha}}}{\frac{\boldsymbol{\sigma}^4_{\boldsymbol{\alpha}_1}}{n_1(n_1-1)} + \frac{\boldsymbol{\sigma}^4_{\boldsymbol{\alpha}_2}}{n_2(n_2-1)}}.$$

The value of `t` obtained from the data is then compared with the corresponding value found in the respective significance table, namely assuming a 95% confidence level. (In our case we have resorted to the tables incorporated into "The R Project for Statistical Computing", https://www.r-project.org)

### The Mann-Whitney-Wilcoxon statistical test

The Mann-Whitney-Wilcoxon test is a rank-based fully nonparametric test whose null hypothesis assumes that the two sets of observations come from the same population whereas the alternative hypothesis states that one of the sets has larger values than the other one. Specifically, assuming that the elements are independent, it is possible to find which observation has the largest value and check the null hypothesis which states the probability that an observation belongs to a first set exceeding an observation from a second one is the same as the probability that an observation from the second set exceeds an observation from the first set and the alternative hypothesis states this probability differs from a set of observations to the other. In practice, the test works as follows: we merge the two sets of values and rank the elements of this new set in ascending order. Then, we perform two sums, $T_1$ and $T_2$, of the rank values conditioned to the group the ranked ($\alpha$) value originally belongs to. From the values of the $T_i$ sums, we define the variables

$$U_i = n_1 n_2 + \frac{n_i(n_i+1)}{2} - T_i, \qquad i = 1, 2.$$

After finding the minimal value, $U_{\min} \equiv \min(U_1, U_2)$, we compare it with the corresponding value found in the respective significance table.

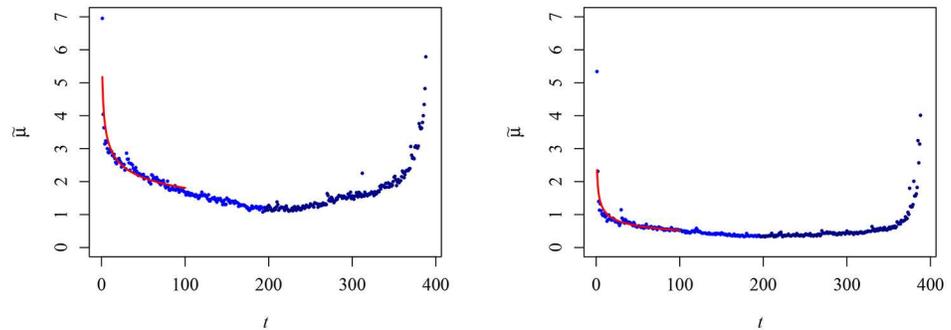

**Fig 1. Average trading volume, $\tilde{\mu}(t;d)$ v intraday time (minutes).** The left panel is for 2S06 and the right panel for 1S13. In both cases we use different tones of blue to represent the morning and evening part of the sessions. The red line represent the numerical fit obtained for Eq (9). In both panels the volume is divided by $10^4$.

doi:10.1371/journal.pone.0165057.g001

## Results and Discussion

### Time dependent statistics

**Mean trading volume.** We naturally start by looking at the well-known ∪-profile of the average of the trading volume, $\mu$, as defined by Eqs (1) and (5)-left. In Fig 1, $\tilde{\mu}(s,t)$ displays its well-known profile with significantly larger values in the beginning and in the end of the trading session; additionally, it is possible to identify a smaller peak at $t = 270$ corresponding to 14:00 in local clock time, which is associated with the closing of European markets.

With respect to the beginning of the session, it is ever more common that, aiming at mitigating its impact on financial markets, companies, watchdogs, governments and other institutions disclose (relevant) information, especially news that are likely to be distressing, after the markets are closed. This news become a latent event whose burden can only be transferred to the price when the market opens. From this perspective, we can connect the opening of the session to the introduction of a shock into the system. Recalling the complex ethos of financial markets, we find resemblance of these shocks with events like earthquakes or avalanches in natural and complex systems [32–34]. For these cases, it is known the system relaxes in a crystal-clear nonexponential form which is well described by a power-law. Assuming

$$\tilde{\mu}(t;s) \sim t^{\alpha}, \tag{9}$$

as in other financial aftershock instances [35, 36], we find good agreement between the data and the power-law relaxation with an exponent around 0.3. In analysing a possible evolution of this exponent in time, we verify an evident sigmoidal profile with the inflexion point corresponding to the second semester 2008 (2S08). Averaging the exponent $\alpha$ over the semesters of each branch we obtain an average value $\alpha = 0.29$ on the pre-2S08 side and $\alpha = 0.37$ on the post-2S08 side, which define the values of each of the two orange horizontal lines in Fig 2.

In order to bestow statistical significance on our claim that the two sub-groups of data actually correspond to different statistical populations, we apply two appropriate independent statistical tests to the results of the exponent $\alpha$, namely the Welch's t-Student and the fully non-parametric Mann-Whitney-Wilcoxon (see details in the Materials and Methods section). In both cases our hypothesis is confirmed within a confidence level of 95%; for the former test we obtained $t_{\alpha} = 6.76 > t_{\text{crit}} = 2.1$ whereas for the latter $U_{\alpha} = 0 < U_{\text{crit}} = 20$.

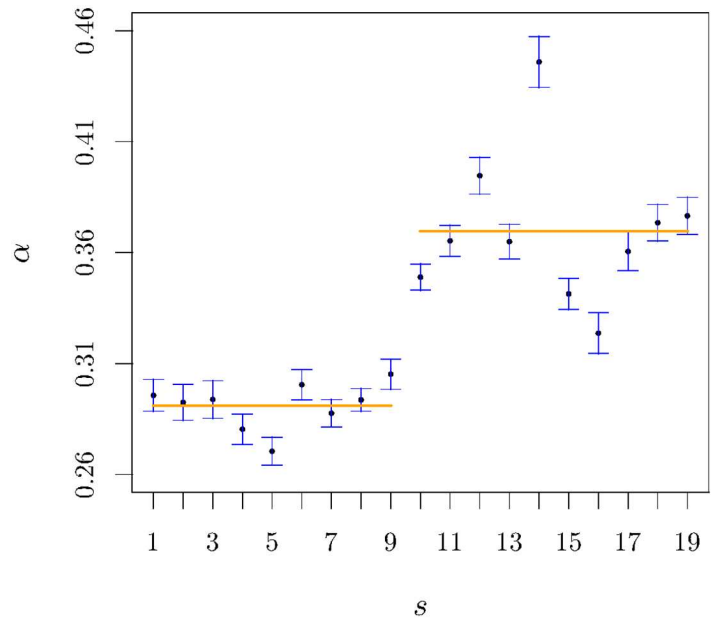

**Fig 2. The symbols represent the exponent α in Eq (9) v semester s.** The solid horizontal lines indicate the value of the exponent averaged over the semesters before and after semester 10. The value of $t$ in the Wech's test of equivalence of the means is equal to 5.56 whereas for critical value with a confidence level of 95% is $t_{crit}$ = 1.86. That difference is corroborated by the Mann-Whitney-Wilcoxon test with a significance of 95% as well.

doi:10.1371/journal.pone.0165057.g002

The market opening and its closing are two completely established events in time. Similarly to the fact that at 9:00 ($t = 0$) the market agents start transferring overnight information to the price until some 'cruise' level of activity is reached, those same agents—especially intraday chartists—prepare themselves to the ringing of the bell at 16:30 ($t = 390$) by increasing the trading volume in the very last part of the session. In many complex system situations and phase transition phenomena in Physics, we have an upsurge of key quantities according to logarithmically modulated power-laws up to some critical value [36]. In the present case—contrarily to those instances, but equivalently to situations involving a deadline [37]—the agents are conscious of the "critical" time, i.e., the closing of the market at 16:30. Moreover, the simple visual analysis of the curves shows there is no such logarithmic modulation of the final part of the trading profile. Nonetheless, it is still sound to fit the last part of the average trading volume curves for a power law proportional to

$$\tilde{\mu}(t;s) \sim |t - 391|^{-\alpha'}, \qquad t > 330. \tag{10}$$

The results of $\alpha'$ as a function of the semester are presented in Fig 3. Although the value of $\alpha'$ increases with $s$, it is visible that the curve is not sigmoidal as it is for the opening exponent. Instead, the evolution of $\alpha'$ is well described by two regions with different slopes. The first regime has a slope equal to 0.046 ± 0.004 and lasts up to 1S07 and for the second regime, which goes from that semester on, the slope increases to 0.1 ± 0.01 (further statistical information is presented on the caption of Fig 3); hence, there is no influence of the semester corresponding to the peak of the sub-prime crisis on the closing exponent exactly as the events that took place in 1S07—and which led to the change in the slope of the $\alpha'(s)$ curve slope—seem to have no

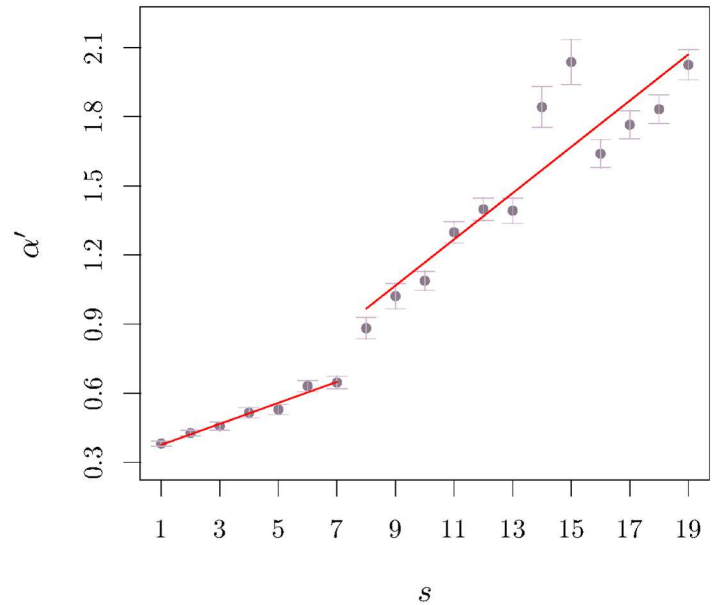

**Fig 3. Closing exponent α′ v semester s.** The symbols represent the exponent α′ in Eq (10) averaged over the companies and the respective deviation bars. The solid lines represent the best fit up to 1S07 with a slope 0.046 ± 0.004 and from that semester we identify a second regime with a slope equal to 0.1 ± 0.01. The correlation coefficient for each of these numerical adjustments is R = 0.97 and R = 0.836, respectively.

doi:10.1371/journal.pone.0165057.g003

relevance in the change of opening relaxation exponent after semester 2S08. Such an observation already hints at a different trading behaviour of the agents during morning and afternoon parts of the session.

Besides the relevance in itself (see Conclusion), this result implies the ∪-shape has been evolving as well. To obtain an overall description of $\mu_i(t; s)$ that could fit every semester without much fretting, we assume a fourth order polynomial

$$\mu_i^{(\text{fit})}(t; s) = c_0 + c_1 t + c_2 t^2 + c_3 t^3 + c_4 t^4, \quad (11)$$

where the coefficients, $c_j$ depend on the company, $i$, and the semester, $s$. Other approaches trying capture the cut-off power-law behaviour at both ends are powered hyperbolic functions. However, Eq (11) has a better pay-off regarding the ratio between the sort of information we are willing to extract and complexity of the numerical adjustment.

In the furtherance of simplicity, the curves $\mu_i^{(\text{fit})}(t; s)$ are effectively obtained after carrying out the change of variables $t \to t/195 - 1$ so that the intraday time gets defined in the interval (−1, 1). Particularly, we focus on the computation of the concavity of each curve $\mu_i(t; s)$,

$$\mathcal{C}_i(s) \equiv \frac{1}{T}\sum_{t=1}^{T} \frac{d^2 \mu_i^{(\text{fit})}(t; s)}{dt^2}. \quad (12)$$

Our computation is presented in Fig 4 for $\langle \mathcal{C}(s) \rangle$ and shows that globally the concavity is in fact diminishing and approaching flatness.

Taking into consideration our observations in the Introduction, we verify that intraday traders apart—who do not carry positions overnight—the first and the last part of the session is influenced by different factors, a characteristic that extends to the evolution of the shape of the

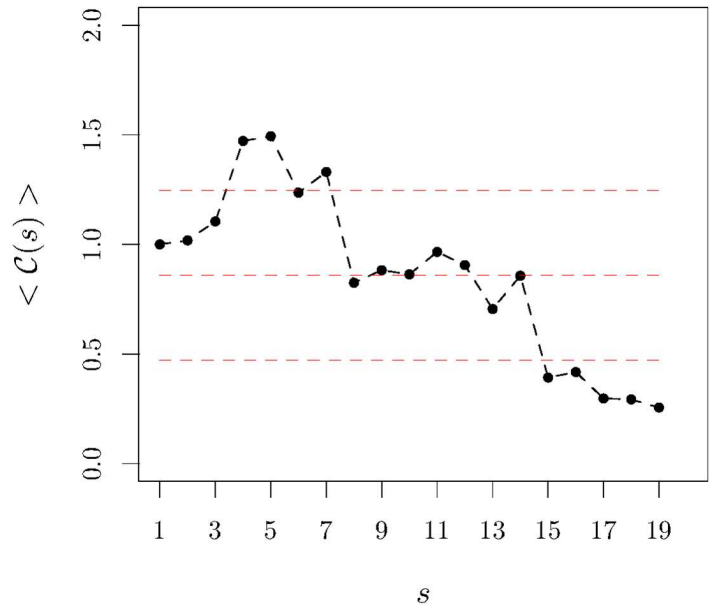

**Fig 4. Concavity of the average traded volume v semester s.** The black dots—with the black line as a guide to the eye—represent the values of the concavity averaged over companies normalised by the value of $\langle \mathcal{C}(1) \rangle$; the horizontal lines signal the average value (computed over all companies and semesters) and the upper and lower error limits. It is clear the concavity of the average traded volume has diminished.

doi:10.1371/journal.pone.0165057.g004

average trading volume curve. Nonetheless, we can estimate the difference between the first and the second half of the session by computing the symmetry of the curve,

$$\mathcal{S}_i(s) \equiv \frac{1}{T}\left[\sum_{t=T/2+1}^{T}\mu_i^{(\mathrm{fit})}(t;s) - \sum_{t=1}^{T/2}\mu_i^{(\mathrm{fit})}(t;s)\right]. \quad (13)$$

The result of that analysis shows the trading profile is more often afternoon tilted, which means the market tends to have the larger stake of its activity in the second half of the session. Such bias is likely to be introduced by longer-term agents who tend to let the markets calm down from its early excitement to open or reinforce their positions. The curve $\langle \mathcal{S}(s) \rangle$ that we introduce in Fig 5 does not show any evident trend aside from the strong peak at semester 11, i.e., after semester 2S08; there is another peak at semester 3, within error bars though. In 2S08, the DJIA index does not evidence significant moves as well as the erstwhile 6-month period. This means, that only further (future) plummetings can provide us with a minimal number of episodes that will allow understanding whether the difference between the morning and the afternoon parts of the trading session relates to large price slumps.

To conclude this part, we pitch at comprehending whether the level of activity in the semester has any relation with the shape of the ∪, namely its concavity. We quantify the activity in the semester by computing the average daily trading volume in the semester

$$\mathcal{V}_i(s) \equiv \frac{1}{N_D}\sum_{d=d_s}^{d_e}\sum_{t=t_i}^{t_f}v_i(d,t;s) = \sum_{t=t_i}^{t_f}\mu_i(t;s) \quad (14)$$

where $t_i$ and $t_f$ represent the first and the last minute of a given day $d$ belonging to semester $s$.

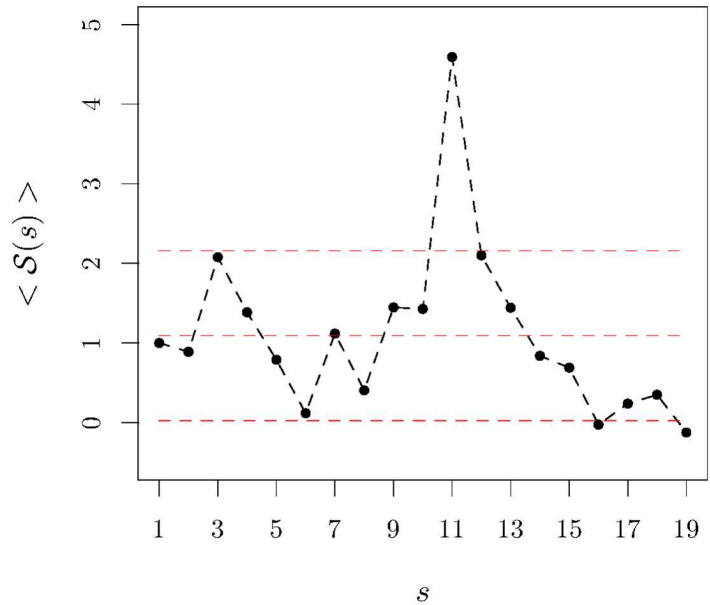

**Fig 5. Symmetry of the traded volume $v$ semester $s$.** The black dots—with the black line as a guide to the eye—represent the values of the mean symmetry normalised by the value of $\langle \mathcal{S}(1) \rangle$; the horizontal lines signal the average value (computed over all companies and semesters) and the upper and lower error limits.

doi:10.1371/journal.pone.0165057.g005

From Eqs (11) and (12), we expect a dependence close to a linear relation between $\mathcal{V}_i(s)$ and $\mathcal{C}_i(s)$ if the coefficient $c_4$ is the most relevant coefficient for the value $\mu_i^{(fit)}(t;s)$. Explicitly, approximating the sums by integrations in the formulae of $\mathcal{V}_i(s)$ and $\mathcal{C}_i(s)$ we have

$$\mathcal{V}_i(s) = 2\,c_0 + \frac{2\,c_2}{3} + \frac{2\,c_4}{5},$$

and

$$\mathcal{C}_i(s) = 2\,c_2 + 4\,c_4,$$

so that with leading $c_4$ it yields,

$$\mathcal{C}_i(s) = -20\,c_0 - \frac{14}{3}\,c_2 + 10\,\mathcal{V}_i(s). \tag{15}$$

That relation is illustrated in Fig 6 with statistical significance. Apart from the small initial values, the red line we obtain by performing a linear regression has a slope equal to 9.45 ± 1.02, a value that is in agreement with our last relation. In other words, the changes in the level of activity in a stock are mainly governed by redefining the intraday shape and not simply by shifting up the curve.

Along the same lines, we test whether $\mathcal{C}_i(s)$—or the average daily trading volume because of their Eq (15)—as well as $\mathcal{S}_i(s)$ have a direct relation with other financial quantities such as the (annualised) volatility, $\Sigma_i(s)$

$$\Sigma_i(s) \equiv \sqrt{\frac{252}{N_D}\sum_{d=d_s}^{d_e}\left[\frac{1}{2}\left(\ln\frac{H(d)}{L(d)}\right)^2 - (2\ln 2 - 1)\left(\ln\frac{C(d)}{O(d)}\right)^2\right]}. \tag{16}$$

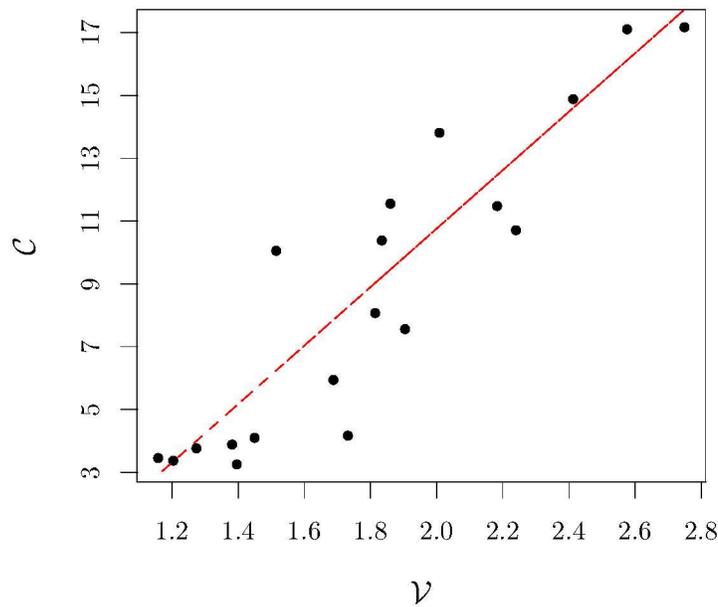

**Fig 6. Relation between concavity and the average daily trading volume for AT&T Inc.** The dots represent the empirical data and the line, which has a slope equal to 9.45 ± 1.02, shows there is a good agreement between our linear relation and the data (the correlation coefficient $R$ of this fit is equal to 0.85). Both quantities are scaled by a factor $10^5$.

doi:10.1371/journal.pone.0165057.g006

with $H(d)$, $L(d)$, $C(d)$ and $O(d)$ representing the highest, lowest, closing and opening price, respectively, and the price variation

$$r_i(s) = 100 \, \frac{S_i(d_e) - S_i(d_s)}{S_i(d_e)}. \tag{17}$$

As regards the concativity—or similarly $\mathcal{V}_i(s)$ —, the putative relation is based on assuming the MDH significant, *i.e.*, if the trading volume and the volatility have same origin, then we expect a quantitative relation between $\mathcal{C}_i(s)$ and $\Sigma_i(s)$ as well. Likewise, if "it takes volume to make the price move", then it should be possible to observe a relation between $\mathcal{C}_i(s)$ and $r_i(s)$ or its absolute value. However, after an analysis over the entire interval, we are unable to find a clear relation between the concavity(symmetry) and the volatility or the price variation. It is worth mentioning that although it has not been possible to grasp any monotonic behaviour, we perceive that for the semesters with larger turbulence the difference between the morning and the afternoon parabolic coefficient shoots up.

**Variance.** After describing the intraday behaviour of $\tilde{\mu}(t)$, we move on to assess the intraday behaviour of the variance. In Fig 7, we show $\widetilde{\sigma^2}(t)$ for semester 7.

The analysis over the entire set of semesters allowed us to understand the ∪-shape evinced by $\widetilde{\sigma^2}(t;s)$ changes in time with a clear asymmetry between the morning and the afternoon parts of the trading session. With these results, we separate out the statistics of the morning from that of the afternoon and perform scatter plots of $\widetilde{\sigma^2}(t;s)$ versus $\tilde{\mu}(t;s)$ that we fit for a parabola, as shown in Fig 8. The results of the numerical adjustments allow us to spot that in frenzied periods (semesters) the difference between the morning and the afternoon parabolic coefficients augments as well as the trading volume dispersion, particularly when $\tilde{\mu}(t;s)$ is

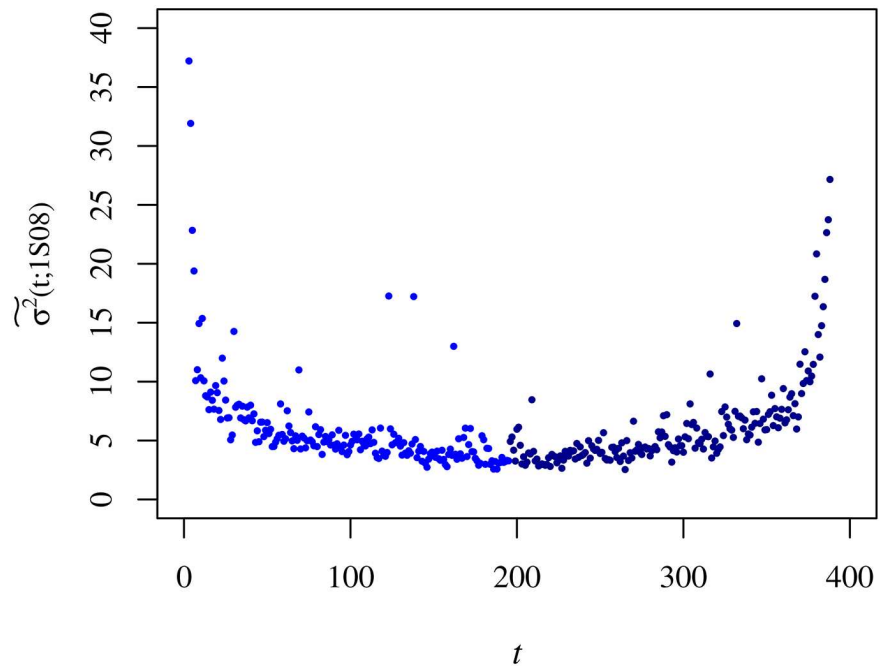

**Fig 7. Variance of the trading volume, $\widetilde{\sigma^2}(t; 1S08)$ v t.** The curve corresponds to semester 7 (1S08) and it was normalised by a factor $10^8$. We use different tones of blue to turn the distinction between the morning and afternoon parts of the session easier.

doi:10.1371/journal.pone.0165057.g007

large. In other words, the trading volume fluctuates less in the afternoon (in average value units).

**Skewness.** In respect of the skewness, it has a ⌒-shape, as visible in Fig 9, with positive values at all times. Due to the little concavity of the curves $\zeta_i(t; s)$, we fit both sides for straight lines and plot each value of the adjustment slope as a function of the semester. Although their absolute values are small—around $1.6 \times 10^{-3} \pm 4 \times 10^{-4}$ when we average over both branches—they are relevant. Moreover, we verify that the difference between them is rather constant as the standard deviation of the difference of the slopes is around 25% of the average value.

We also compare the relation between $\tilde{\zeta}(t; s)$ and $\tilde{\mu}(t; s)$, for which we find an (expected) inversely proportional relation, i.e., the larger the average, the shorter the tail on the righ-hand side of $p(v)$. Using linear fits (for the sake of simplicity), we check that the relation is almost the same for the morning and the afternoon. Furthermore, those relations are pretty constant (within expectable fluctuation) up to semester 16 (1S12) with a strong change, namely an augment in the value of the slope, afterwards.

**Kurtosis.** To conclude our account on the intra-day individual statistical properties of the trading volume, we present the results of our analysis on the intraday profile of the kurtosis, $\tilde{\kappa}(t; s)$, showing that it has a leptokurtic ⌒-profile during all the trading session, as seen in Fig 10. That corresponds to a striking difference when we compare it with the profile which was found for the kurtosis of the price fluctuations. Explicitly, the kurtosis starts with a very large value and it lessens until half of the morning period, whence it partially rallies until the start of the afternoon period of trading (12:45-13:00). From that time onwards, the kurtosis lessens. Additionally, the kurtosis often has spikes around the morning-afternoon transition and in the two last minutes of the trading session.

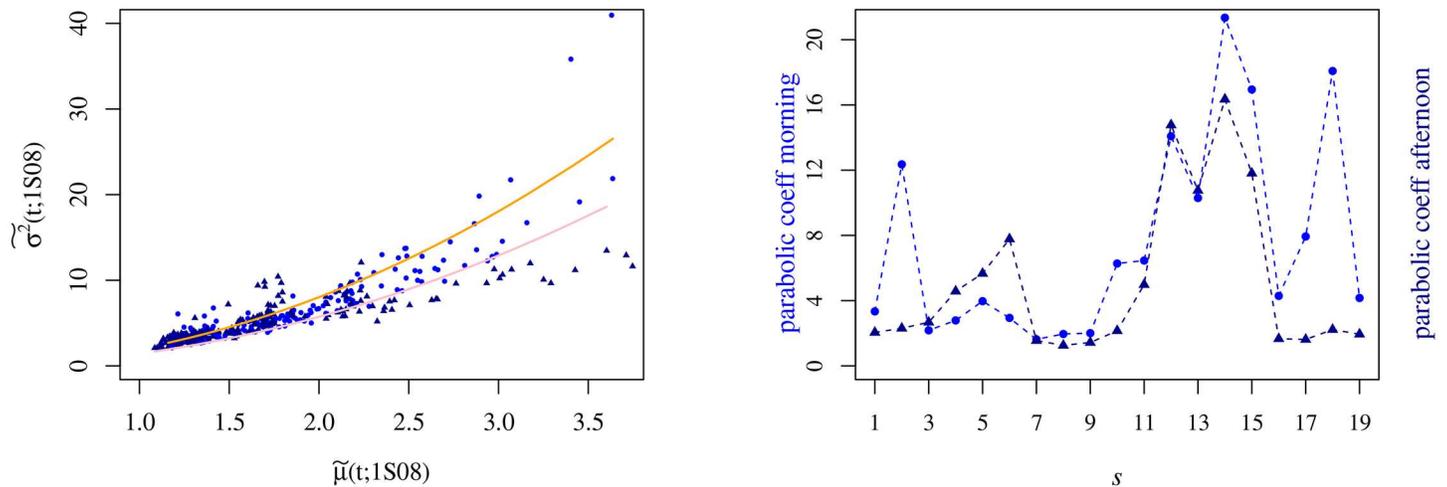

**Fig 8. Relation between the variance and the average of the trading volume in 1S08.** For both panels the ultramarine blue dots represent the morning values and the navy blue triangles the afternoon ones. Left Panel: Variance v mean trading volume averaged over all the companies for $s = 9$. The values of $\tilde{\sigma}^2(t;s)$ and $\tilde{\mu}(t;s)$ are scaled by $10^8$ and $10^4$, respectively. The orange and the pink lines represent the parabolic fit for curves of the morning and the afternoon periods yielding the following parameters: 2.01 ± 0.06 and 1.43 ± 0.04, respectively. Right panel: Value of the parabolic coefficient of the morning and the afternoon parts of the trading session as a function of the semester.

doi:10.1371/journal.pone.0165057.g008

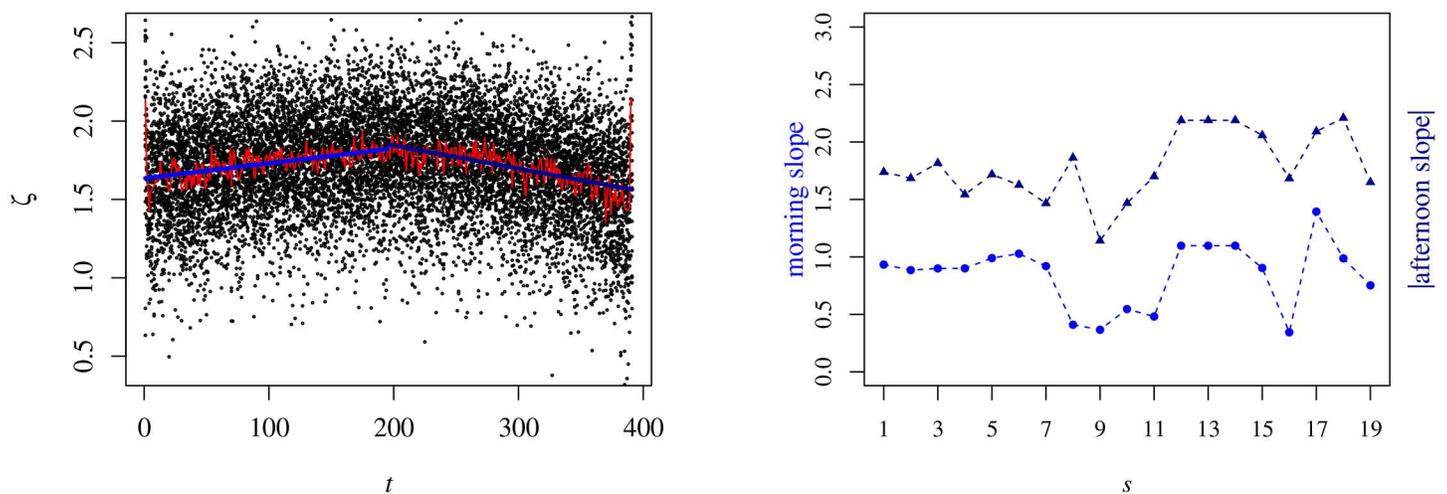

**Fig 9. Intraday profile of the skewness during 1S04.** Lef panel: The black dots represent the values of $\zeta_i(t; 1S04)$, the red line $\tilde{\zeta}(t; 1S04)$ and the ultramarine and navy blue lines corresponds to the linear fit of the red curve for the morning and afternoon parts of the session whose slopes are $10^{-3} \pm 10^{-4}$ and $1.5 \times 10^{-3} \pm 10^{-4}$, respectively. Right panel: The values of the slopes of the fittings as a function of the semester. The ultramarine blue dots represent the morning values ofthe slope and the navy blue triangles the absolute value of the slope in the afternoon part of the trading session.

doi:10.1371/journal.pone.0165057.g009

The behaviour of the kurtosis can be further explored looking into its relaxation after the market opens and before it closes. To that, we understand that each case is well described by

$$\langle \kappa(t;s) \rangle \sim t^{-\beta_m} \quad \text{if} \quad t < 100$$

$$\langle \kappa(t;s) \rangle = A - B(t - 290)^{\beta_a} \quad \text{if} \quad t > 290, \quad (18)$$

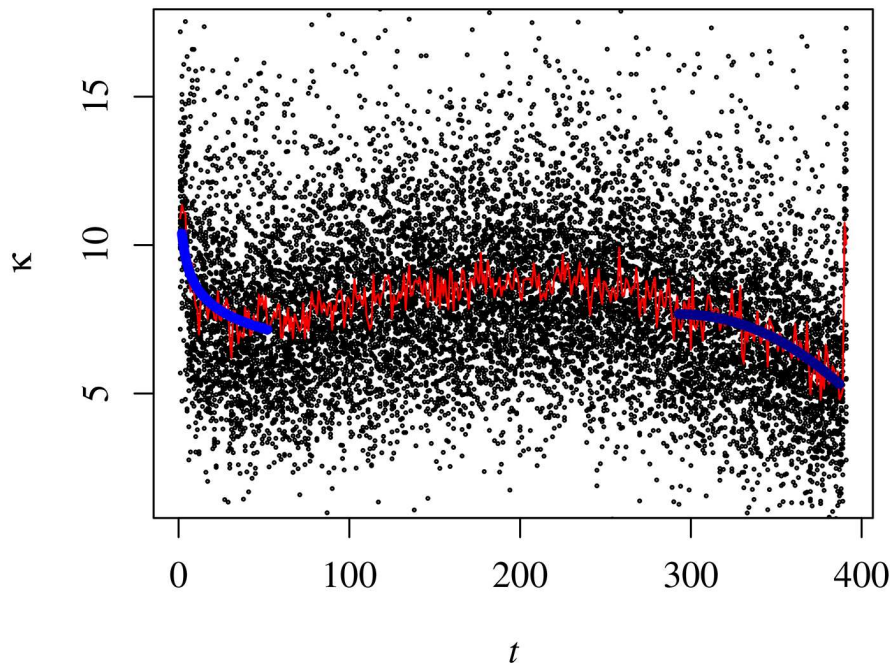

**Fig 10. Intraday profile of the Kurtosis during 1S04.** The black dots represent the values of $\kappa_i(t; 1S04)$ and the red line $\tilde{\kappa}(t; 1S04)$. The ultramarine and navy blue lines represent the adjustments given by the first and second formulae in Eq (18), respectively. In this specific case the parameters are $\beta_m = 0.11 \pm 0.02$ and $\beta_a = 2.16 \pm 0.03$.

doi:10.1371/journal.pone.0165057.g010

where $A$, $B$, $\beta_m$ and $\beta_a$ depend implicitly on the semester. In Fig 11, we show the evolution of the exponents across the semesters.

With respect to $\beta_m$, we cannot find any specific behaviour besides a consistent increase in the last six semesters, which is noteworthy though. For the other exponent related to the afternoon, $\beta_a$, there is an apparent saw-like evolution. Comparing both values, we verify that, on average, $\beta_a$ is more than ten times as large as $\beta_m$. Explicitly, we have approximate average values equal to 0.13 for the former and 1.7 for the latter.

Still regarding the nonstationarity of the kurtosis, from the scatter plots $\tilde{\kappa}(t; s)$ v $\tilde{\mu}(t; s)$ (see Fig 12), there emerges a striking difference between the morning and the afternoon parts of the session. While in the morning period there is a visible parabolic dependence of the kurtosis on the average trading volume for all the semesters, the second order coefficient of the fitting of the afternoon curves vanishes from "sub-prime" semester 10 (2S08) onwards. For the remaining fitting coefficients, it is noticeable an increase of the linear term and a decrease of the constant coefficient.

## Cross-sectional statistics

In the previous section, we have introduced the results concerning the statistics of the trading volume across intraday time (conditioned to semester $s$) and presented them carrying out averages over trading dates. Another way of having a look at the problem is to take into consideration the fact that we have a set of companies that do not behave equivalently. For instance, previous results presented in [27, 38] showed that the distribution of price fluctuations among stocks gets more Gaussian when stock markets have large collective (index) fluctuations. In the

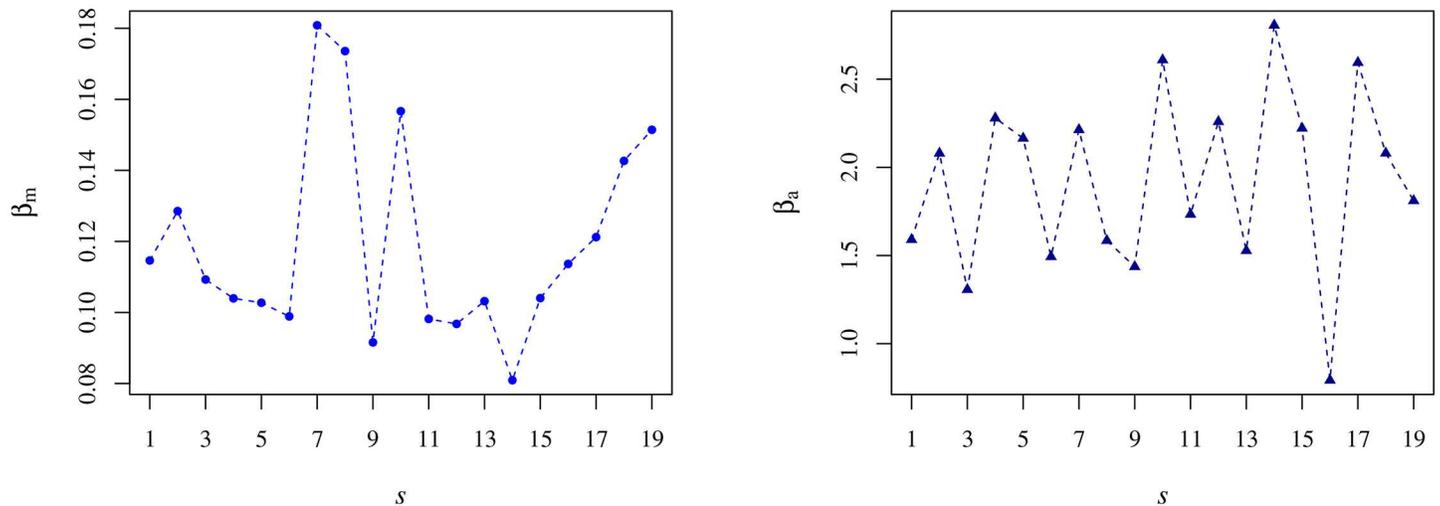

**Fig 11. Relaxation exponents of the kurtosis in Eq (18) as a function of the semester.** The left panel depicts the evolution of the morning exponent, $\beta_m$, and the right panel the same but for the afternoon exponent, $\beta_a$. Both exponents differ from each other by a factor 10.

doi:10.1371/journal.pone.0165057.g011

current section, we analyse the cross-sectional statistics of trading volume that will help us cast light on the behaviour of the companies as a group.

Because we are interested in the overall results and $\tilde{\mu}(t;s) = \hat{\mu}(t;s)$, we go directly to compute the variance. Nevertheless, it is appealing to analyse the statistics of the companies across the days (belonging to a given semester) in order to understand their relative behaviour.

**Variance.** Using Eq (6), we observe the cross-sectional variance, $\hat{\sigma}^2(t;s)$, has a ∪-shape similar to both $\tilde{\mu}(t;s)$ and $\tilde{\sigma}^2(t;s)$ obtained for the individual statistics. Comparing $\hat{\sigma}^2(t;s)$ with $\tilde{\sigma}^2(t;s)$, we attest to the fact that the former is significantly smaller than the latter at all times; in other words, the fluctuations in the trading volume of a company are larger than the dispersion of that quantity among stocks (see Fig 13). Contrarily to what was observed for the price fluctuations [27], we notice that the dispersion tends to increase as the trading session evolves seeing that the variance ratio $\tilde{\sigma}^2/\hat{\sigma}^2$ (slowly) decreases. Nonetheless, in the very first and last minutes of the trading session we see an increase(decrease) of the variances ratio (dispersion).

Since the cross-sectional variance has revealed a ∪-shape, we can analyse its concavity and symmetry by applying formulae equivalent to Eqs (12) and (13). These results are presented on a semester-by-semester basis in Fig 14, where it is visible a strong increase of the two quantities between semesters 11 and 14 and to which we shall be back when discussing the cross-sectional kurtosis. Fig 14 shows a conspicuous drop in the variance ratio. Concomitantly, we have a strong increase in the difference between the morning and afternoon variance values. It is worth mentioning that differently to the case of the mean trading volume, the value of $\mathcal{S}$ for either of the variances is negative.

**Skewness and Kurtosis.** With respect to the higher order cumulants their intraday profiles are rather noisy, a fact that is certainly due to the size of our set. Nonetheless, we can point out some features that are sufficiently robust, namely: for the *cross-sectional skewness* we grasp that the trading volume is regularly positive at all times and $\hat{\zeta}(t;s)$ has a ⌒-profile, as happens for

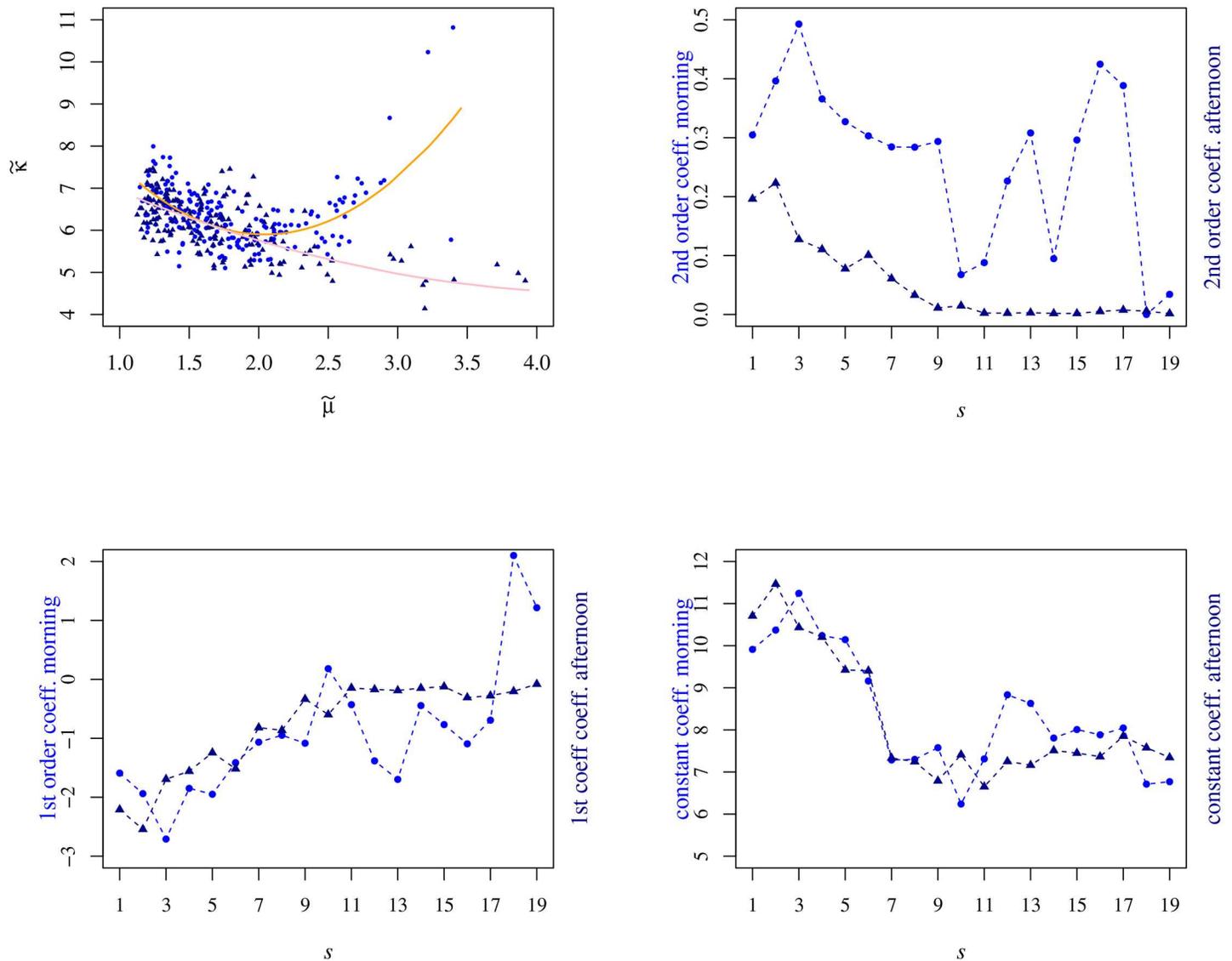

**Fig 12. Relation between the kurtosis and the mean trading volume.** In the upper left panel we present the scatter plot $\tilde{\kappa}$ v $\tilde{\mu}$ for 1S06 where the lines correspond to the fits with a second order polynomial for each half of the trading session. For the morning orange line fit we have $\tilde{\kappa} = (1.5 \pm 0.1)\tilde{\mu}^2 - (6.1 \pm 0.5)\tilde{\mu} + (12.2 \pm 0.5)$ whereas for the afternoon pink line fit $\tilde{\kappa} = (0.19 \pm 0.08)\tilde{\mu}^2 - (1.8 \pm 0.3)\tilde{\mu} + (8.5 \pm 0.4)$. The remaining panels present (in counterclockwise direction) the evolution of the second, the first and the constant (order) polynomial coefficients.

doi:10.1371/journal.pone.0165057.g012

the individual stocks; however, in applying linear fits to each part of the profile, we determine slopes of the same order of magnitude of the error.

For the *kurtosis*, our computations have indicated a regular decrease of $\hat{\kappa}(t;s)$ in the first hour of trading and after that, a relatively constant profile (bearing in mind the fluctuations). In order to have a clearer picture, we compute the average cross-sectional kurtosis for the trading session (constrained to $t > 60$) and plot the results for each semester as shown in Fig 15. From that figure we understand that, on average, the cross-sectional distribution of the trading volume is close to mesokurticity (actually slightly above). The behaviour of the kurtosis

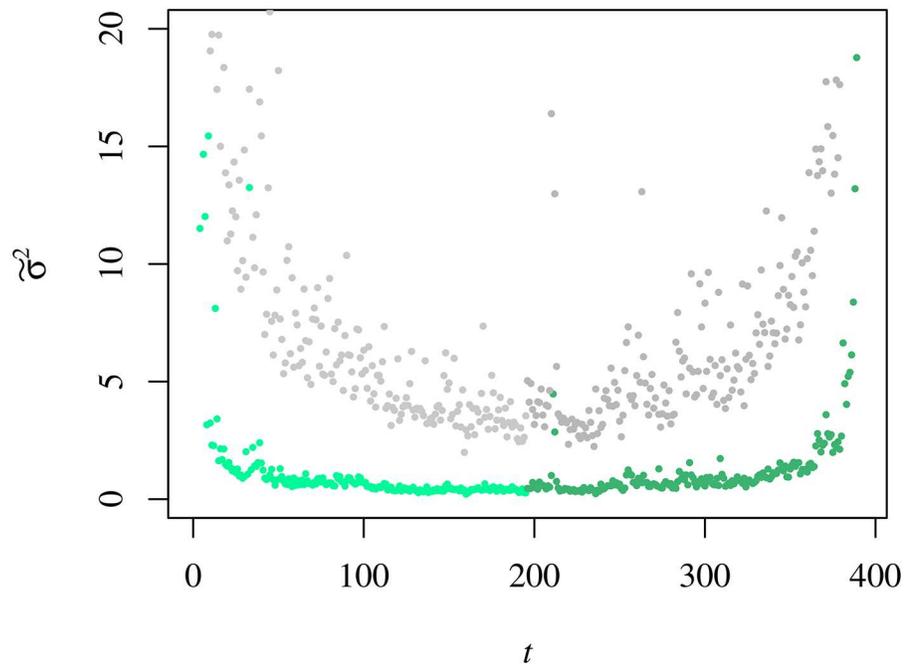

**Fig 13. Cross-sectional variance $v$ intra-day time in 2S08.** As employed for the individual analysis we use the lighter(darker) tone of green for the morning(afternoon) part of the curves $\hat{\sigma}^2(t; 2S08)$. For mere comparative purposes we present the individual analysis $\tilde{\sigma}^2(t; 2S08)$ in grey.

doi:10.1371/journal.pone.0165057.g013

significantly augments between semester 11 and semester 16, which includes the aforementioned period for which both the concavity and the symmetry of the cross-sectional variance boosted. In the first three semesters of that spell, General Motors was not traded for the reasons we have already explained. Yet, the larger values of the kurtosis occur when GM.NY trading had resumed. Note that, in this same period,—which concurs with the period showing stronger fluctuations of the relaxation exponent $\alpha$ (see Fig 2)—we find a noteworthy increase of the ratio between $\sqrt{\tilde{\sigma}^2}$ and $\tilde{\mu}$ (see Fig 8). In that span, the DJIA rocketed from 9034.69 to 12217.86 points, i.e., a rally of 35%, with a mininum value of 6626.94 on the 6th March 2009 and fetching a maximum of 12810 points on the 29th April 2011. In order words, the period spanning between semester 11 and 16 is typified by a hefty rise in the index.

With the goal of comprehending more effectively the typical behaviour of the kurtosis among stocks, we exclude the anomalous semesters (between $s = 11$ and $s = 16$) and averaged $\hat{\kappa}(t; s)$ over the remaining ones; that yields the all-inclusive average kurtosis $\kappa(t; s)$. The intra-day profile of $\kappa(t)$ is put on display in Fig 16. That profile runs *grosso modo* along with the ⌢-profile of the kurtosis of the trading volume of individual stocks, $\tilde{\kappa}(t; s)$, but with smaller values. Therefrom, we see that in respect of the trading volume, the DJIA set of stocks starts strongly non-Gaussian, relaxing to Gaussianity after about 1 hour of trading and remaining around there for the next hour; from that time on, we observe a surge in the kurtosis until circa 13:00h. Afterwards, $\kappa(t)$ relaxes to Gaussian values with a rebound in the very last minutes — which we associate with eleventh hour trading events —, but much smaller than the initial values.

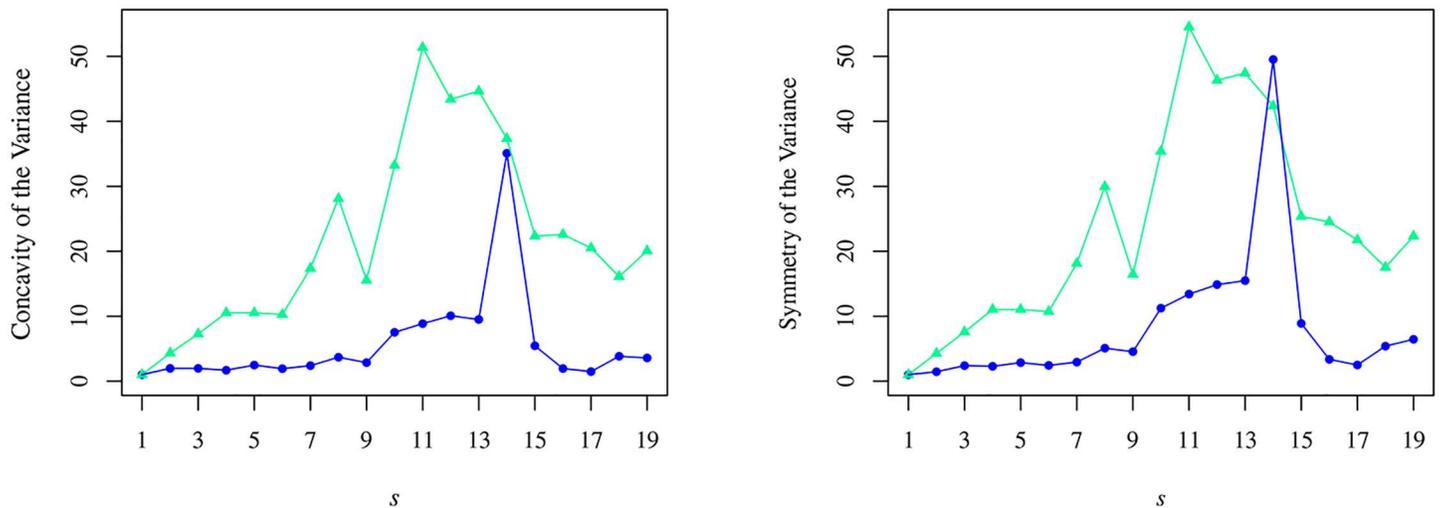

**Fig 14. Individual and Cross-sectional Concavity (left) and Symmetry (right) v semester *s*.** In both panels, the green triangles represent the cross-sectional variance, $\hat{\sigma}^2$, and the blue dots stand for the individual variance, $\tilde{\sigma}^2$. The values in the abscissae are normalised by the value in the first semester.

doi:10.1371/journal.pone.0165057.g014

## Conclusion

The present work has aimed at extending the analysis of the celebrated ∪-shape of the trading volume in financial markets. To that, we have used 1-minute sampling rate data of blue chip Dow Jones Industrial Average equities spanning the years between 2003 and 2014. Trading volume is widely viewed as a proxy for information in a financial market, a feature that utterly justifies the relevance of our work. We have done so twofold; in first place, by studying the intraday behaviour—specifically the profile—of the cumulants of the trading volume up to order four and then by looking to the nonstationary features of those profiles. Due to the latter analysis, we have split our data into contiguous patches of six months. Additionally, we have considered two kinds of approach: *i)* performing the statistics for each company and then averaging over them (individual analysis) and *ii)* carrying out the computation of the statistical moments of the trading volume over the companies (cross-sectional analysis). In both cases, we have confirmed that the trading volume is significantly nonstationary, at least for the period we have studied.

For the individual analysis, using an analogy between the decrease in the average trading volume after the opening of the market and the relaxation observed after earthquakes and avalanches in complex systems, we have adjusted the initial part of the average trading volume curve with a power-law and verified that the market changed its early trading behaviour after the second semester of 2008, i.e., the climax of the subprime crises. Explictly, we have found that $\tilde{\mu}(t < 100; s)$ was decaying more slowly before 2S08 than it started taking place after that semester. That change—which we have corroborated employing different statistical tests—translates into the following figures: before 2S08 the market was taking 10 minutes to reach an average trading volume that was half of its value in the first minute whereas after 2S08 the time needed to get half of the opening trading volume value slid to 6 minutes, a drop of 40%. Within the period our data accounts for, the modifications with major potential impact on trading we have been able to find correspond to the withdrawal of the "uptick rule" in 2007 [39] and its reintroduction upon revision in 2010 [40]. Looking at the figures of $\alpha$, we have recognised that the value of the exponent did not change until 2S08; equivalently, after the reinstitution of

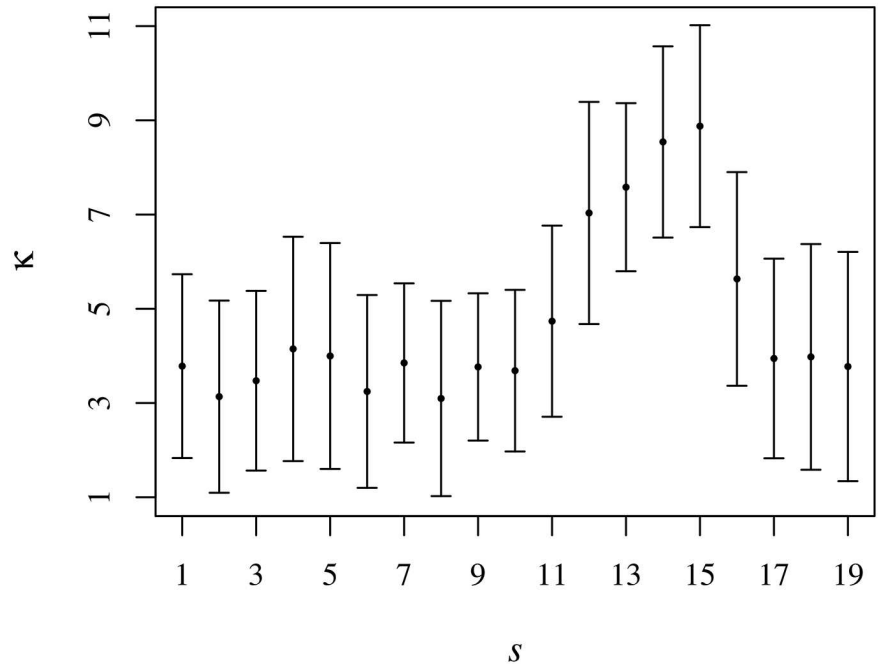

**Fig 15. Intraday cross-sectional kurtosis $v$ semester $s$.** The symbols (and the bars) represent $\hat{\kappa}(t;s)$ averaged over intraday time with $t > 60$.

doi:10.1371/journal.pone.0165057.g015

those rules, the relaxation exponent has kept its post-2S08 value. Thence, we have assigned the change in the value of $\alpha$ to an effective modification in the trading conduct by agents after the crisis. That claim is bolstered by the analysis of the behaviour towards the ringing of the bell at 16:30; by adjusting the last part of the $\tilde{\mu}(t;s)$ with a power-law we have found a monotonic increase in the value of the exponent followed by a crossover in the first semester of 2007. In other words, the changes in the rules for short-selling are likely to have affected the closing of the session and had negligible impact on the opening.

The study made on price fluctuations has shown that their intra-day time dependent standard deviation decays in the first part of the session similarly to a power-law with an exponent around 0.3. The closeness of that exponent with the value we have computed for the relaxation of $\tilde{\mu}(t;s)$ hints at a relation between volatility and trading volume that would fit for the Mixture of Distribution Hypothesis. Accordingly, we expect that ongoing work over the nonstationarity intraday relation between the trading volume and the volatility will shed light on this matter [41]. To bear out the changing of the trading profile, we have assessed the concavity of the ∪-shape and verified that this quantity lessened in time, which sustains our assertion that the well-known intraday profile is ever more turning into a ⊔-shape. In addition, we have verified there is an asymmetry in the curve of the average trading volume as a function of the intraday time; yet, in this case it has been impossible to specify any trend whatsoever. We have also confirmed that the form of the curve $\tilde{\mu}(t;s)$ plays a major role in the increase(decrease) of the trading activity, i.e., the activity in the market is not just related to shift upwards(downwards) of the trading volume profile, but it mainly affects its specific shape instead. We have endeavoured to connect the properties of the average trading volume curve with quantities like the price fluctuation and the volatility in the respective period; however, we could not find any clear relation between these quantities. This fact is at odds with the results on the relaxation of the

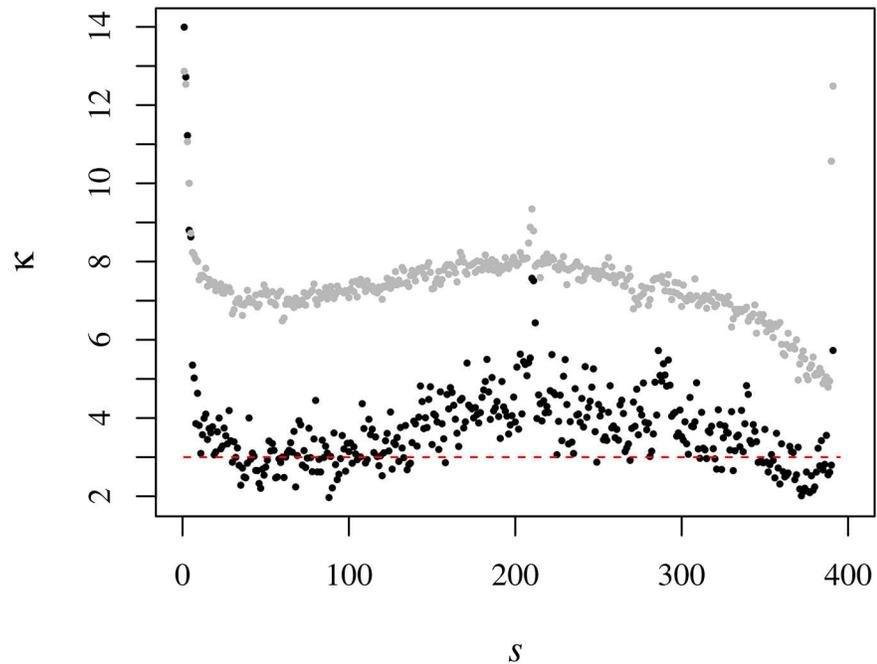

**Fig 16. Intraday kurtosis $v$ time $t$.** The black dots represent $\hat{\kappa}(t;s)$ averaged over the semesters (excluding those in the interval between $s = 11$ and $s = 16$). For mere comparison purposes, we present $\tilde{\kappa}(t;s)$ using the grey symbols as well. The dashed red line represents the value of the kurtosis for Gaussian distributed variables.

doi:10.1371/journal.pone.0165057.g016

average trading volume, which emphasises the need for additional analysis on the relation between $v$ and the volatility.

The skewness presents a ⌢ profile; nonetheless, the concavity of that profile is so small that we have opted to fit that profile linearly. The adjustment substantiated the existence of different behaviour in the morning and the afternoon trading periods as the average slope of the latter is 150% as large as the average slope of the former. Nevertheless, the most vivid expression that the trading in the morning and in the afternoon are governed by different mechanisms comes from the kurtosis, for which we have computed a ⌣-like profile, where the afternoon part of the curve concurs with the bending down part of the profile. Interestingly, the intraday shape of the kurtosis is able to identify the so-called "lunch effect", for which an increase in the activity after many agents resuming their trading activities is expected. Yet, our analysis has showed that its impact goes beyond the mere return to trading. Note that apart some from a very located spikes, the effect does not appear in $\tilde{\mu}(t)$ nor $\tilde{\sigma}^2(t)$, but gets utterly clear for the kurtosis instead. Because the kurtosis—sometimes referred to 'kurtosis excess' as well—is a way of measuring the distance to the *Normal* distribution, we can interpret it as a measure of the "surprise" incorporated to the distribution. Along this view, we have comprehended that at the beginning of the trading session, there is a large uncertainty in the value of the trading volume that is kept up with a large value of the surprise as well. That relation is in contrast with the last part of the session, where the hiking up of the fluctuations is not that surprising and the kurtosis decreases. We read these facts as though in some sense, part of the information that was transferred into the stock price in the first half of the morning is assimilated by the market during the session—namely on the asset price—whereas the last part of the session is dominated by corrections to the prices—with no significant feeding in of relevant/impacting information—that goes along with the closure of intraday trader's positions. Pairing the kurtosis with the average of the trading volume, we have noticed that in the morning their relation is robustly parabolic whereas in the afternoon period we verified that the parabolic (second order) coefficient dwindled until the crisis semester 2S08 and effectively vanished from that semester onwards. Once again, we have confirmed the three lessons it is possible to learn from our results: *i)* Besides the well-known intraday seasonality of the ∪-shape, the trading session has clearly different morning and afternoon dynamics; *ii)* the different relation between cumulants, especially the kurtosis and the average, for each part of the trading session indicate a change in the form of the distribution beyond a simple modification of the values of the parameters, an effect that can be obtained by (un)zeroing dynamical parameters in the set of stochastic equations that are set forth to mimic trading volume time series; and *iii)* those features have evolved, at least, for the last decade with the second semester of 2008—the semester wherein the subprime crisis reached its climax—marking the emergence of changes in the trading behaviour that modified the statistics of trading volume in a significant way.

Inspecting the behaviour of the stocks as elements of a group (cross-sectional statistics), we have understood the same qualitative behaviour we found for the individual variance and the kurtosis. Quantitatively, the values of the cross-sectional analysis are smaller than those obtained for the individual approach. In the first part of the morning, we have observed that both the dispersion and the kurtosis decay; this can be read as follows: mainly due to overnight episodes, there is a different level of activity among the stocks with some of them—eventually linked to relevant information disclosed when the market was closed—having abnormal values of trading. That flustering is then dissipated—i.e., transferred to the stock prices—in the first hour of trading with the trading volumes of the equities relaxing to a group of elements Gaussian distributed. Leading up to the beginning of the afternoon part of the session, the kurtosis increases, a fact that we associate with the dual effect of contrarian strategies or intraday traders willing to secure some profit on the stocks that have shown abnormal activity at the opening. Afterwards, there is a relaxation to the Gaussian and the stocks behave among them as expected—i.e., none of them presents (relative) outstanding trading levels and pointing to an increasing of mean trading value $\mu_i(t;s)$—although the dispersion waxes as the end of the session gets closer. As regards the nonstationary cross-sectional behaviour, we have noticed that the period between 1S09 and 1S11 is marked by a solid increase in the cross-sectional kurtosis, signalling that the 35% rally of the DJIA in that period is likely to be fuelled by stocks which had abnormal trading levels.

That said, contrary to the average price fluctuations [27], we conclude the ∪-shape of the trading volume stems from different reasons depending on branch of the curve: in the morning, it hinges on companies that are heavily traded because of some overnight event that needs to be transferred to the price as soon as the market opens—if only to clampdown on arbitrage opportunities —, whereas the augment in the last part of the session is related to a more general increase in the level of trading among all the stocks.

With these results in hand we can think of further developments, namely: *a)* examine the same intraday statistical profiles of trading volume in other markets, their nonstationarity and measure quantities like the exponent of relaxation $\alpha$ as well as the concavity of $\tilde{\mu}(t;s)$ and $\tilde{\sigma}^2(t;s)$ in order to quantify the impact of the level of liquidity and maturity on those properties —e.g., we have noticed that $\mu(t;s)$ for the two NASDAQ stocks taking part in the DJIA index is clearly larger than the typical values for the NYSE stocks —, *b)* extend the set of companies with the goal of perceiving the influence of the size of a company, namely its capitalisation, in the form of its trading volume seasonalities and *c)* analyse crossed intraday profiles of the trading volume with the volatility and the price fluctuations in the spirit of the so-called leverage effect.

## Acknowledgments

We would like acknowledge Olsen Financial Data and Frédéric Abergel from the Chair of Quantitative Finance of the École CentraleSupélec for having provided us with the data. We also would like to thank M. Naeeni for reading the manuscript.

## Author Contributions

**Conceptualization:** SMDQ.

**Data curation:** MBG.

**Formal analysis:** MBG SMDQ.

**Funding acquisition:** SMDQ.

**Investigation:** MBG SMDQ.

**Methodology:** SMDQ.

**Project administration:** SMDQ.

**Resources:** MBG.

**Software:** MBG.

**Supervision:** SMDQ.

**Validation:** MBG SMDQ.

**Visualization:** MBG.

**Writing – original draft:** MBG.

**Writing – review & editing:** SMDQ.